\documentclass[aps,12pt,preprint,prd]{revtex4}
\usepackage{txfonts}
\usepackage{epsfig,dcolumn,bm}

\begin{document}

\title{$QQ\bar{q}\bar{q}$ Four-quark Bound States in Chiral
SU(3) Quark Model}

\author{M. Zhang$^{1,2}$}\thanks{E-mail: zhangmin@mail.ihep.ac.cn}
\author{H.X. Zhang$^{3}$}
\author{Z.Y. Zhang$^{1}$}
\affiliation{\small
$^1$Institute of High Energy Physics, P.O. Box 918-4, Beijing 100049, P.R. China\footnote{Mailing address.} \\
$^2$Graduate School of the Chinese Academy of Sciences, Beijing,
P.R. China\\
$^3$Capital Medical University School of Biomedical Engineering,
Beijing 100069, P.R. China\\
}


\begin{abstract}
The possibility of $QQ\bar{q}\bar{q}$ heavy-light four-quark bound
states has been analyzed by means of the chiral SU(3) quark model,
where $Q$ is the heavy quark ($c$ or $b$) and $q$ is the light quark
($u$, $d$ or $s$). We obtain a bound state for the
$bb\bar{n}\bar{n}$ configuration with quantum number
$J^{P}=1^{+},I=0$ and for the $cc\bar{n}\bar{n}$ ($J^{P}=1^{+},I=0$)
configuration which is not bound but slightly above the $D^{*}D^{*}$
threshold (n is $u$ or $d$ quark). Meanwhile, we also conclude that
a weakly bound state in $bb\bar{n}\bar{n}$ system can also be found
without considering the chiral quark interactions between the two
light quarks, yet its binding energy is weaker than that with the
chiral quark interactions.
\end{abstract}



\maketitle

\section{Introduction}

     A great interest in studying four-quark (4q) states has been
triggered by the observations of some very narrow resonances,
e.g., $D^{*}_{sJ}(2317)$ by BABAR \cite{bar01}, $D_{sJ}(2460)$ by
CLEO \cite{cleo02}, $X(3872)$ by BELLE \cite{belle03}, and
$Y(4260)$ by BABAR \cite{bar04}. Because almost all of these new
states have the same special properties, which are narrow and
their structures are difficult to be explained as the simple
quark-antiquark pair, much attention has been paid and various
theoretical explanations were proposed. Mostly, these new states
were suggested as good candidates of 4q states \cite{es05}. In
this work we will try to see if there are possibilities for the
existence of the four-quark states with two heavy quarks which is
also a significant problem in the new hadron state studies.

      The possible existence of 4q states composed of two heavy
quarks (anti-quarks) and two light anti-quarks (quarks) has been
suggested and studied in the past few years in different ways
\cite{06,car07,pep08}. For example, the work of Carlson et al.
\cite{car07}
based on the MIT bag model concluded that the state
$bb\bar{u}\bar{d}$ with $J^{P}=1^{+}$ is bound and the
$cc\bar{u}\bar{d}$ is not bound. A different conclusion is
obtained by Pepin et al. \cite{pep08} using a pseudoscalar meson
exchange interaction instead of the chromomagnetic potential, and
their results indicated that such interaction binds the
heavy-light 4q systems both for $Q=b$ and $Q=c$. Therefore,
whether the bound state exists in heavy-light 4q systems is still
an open question.

    In this work, we focus our attention on the heavy-light 4q states
and study their masses in the framework of chiral quark model (CQM).
Since the heavy-light four-quark state contains two light quarks
(anti-quarks) which are regarded as constituent quarks, the
non-perturbative QCD effect should be considered between them. On
the other hand we notice that the chiral quark model \cite{zy9,zy10}
based on the idea that between the chiral symmetry breaking scale
and the confinement scale, QCD may be formulated for the light-quark
sector as an effective theory of constituent quarks interacting
through Goldstone boson modes associated to the spontaneous breaking
of chiral symmetry, and CQM has obtained great success in
reproducing the energies of the baryon ground states, the binding
energy of deuteron, the nucleon-nucleon phase shifts, and the
nucleon-hyperon cross sections \cite{zy10,dai11}. Thus one can
expect that the CQM is an effective approach to study the 4q systems
with light quarks. In this work, the four-quark states with quark
content $cc\bar{s} \bar{s}, cc\bar{n}\bar{n}, bb\bar{s} \bar{s}$ and
$bb\bar{n}\bar{n}$ are studied in the chiral SU(3) quark (CSQ)
model.

The parameters for the light quark pairs are taken from our previous
work, and for the heavy-light pairs they are obtained from fitting
the masses of ($D, D^{*}, D_{s}, D^{*}_{s}, \eta_{c}, J/\psi,
h_{c}(1p)$), and ($B, B^{*}, B_{s}, B^{*}_{s}, \eta_{b},
\Upsilon(1s)$), the details will be introduced in the section II.B.
By using these parameters, we obtain a bound state for
$bb\bar{n}\bar{n}$ configuration and no bound state for
$cc\bar{q}\bar{q}$.

Furthermore, we calculate the masses of $QQ\bar{n}\bar{n}$ without
considering chiral quark interaction (CQI), and compare them with
the masses of $QQ\bar{n}\bar{n}$ considering CQI. Through comparison
we uncover that a bound state in $bb\bar{n}\bar{n}$ system can also
be found without considering CQI, yet this bound state is not as
strong as the bound state with CQI.

     The paper is arranged as follows. In the next section,
the theoretical framework of CSQ model, the determination of
parameters and the wave functions for $QQ\bar{q}\bar{q}$ stats are
briefly introduced. The calculated results are listed in Section
III, where some discussions are made as well. Finally, the summary
is given in Section IV.

\section{Theoretical framework}
\subsection{The model}

In the chiral SU(3) quark (CSQ) model, the Hamiltonian of a
$QQ\bar{q}\bar{q}$ four-quark system can be written as
\begin{eqnarray}
\label{hami6q} H=\sum_{i} m_{i}+\sum_{i} T_{i}-T_{G}+
V_{QQ}(12)+V_{\bar{q}\bar{q}}(34)+\sum_{i=1,2,j=3,4}V_{Q\bar{q}}(ij),
\end{eqnarray}
where $m_{i}$ is the mass of the ith quark, $T_{G}$ is the kinetic
energy operator for the center-of-mass motion.
Since the light quark is treated as constituent quark, and the
non-perturbative QCD effect between them has to be considered,
thus for the light quark pairs, besides the confinement potential
and OGE interaction, the chiral quark interactions should be
included also, but which is unnecessary to be considered for the
heavy-heavy and heavy-light quark pairs. The expressions of the
interactions are written as follows:
\begin{eqnarray}
\left\{\begin{array}{l} V_{QQ}(12)=
V_{QQ}^{conf}(12)+V_{QQ}^{OGE}(12),\\
V_{\bar{q}\bar{q}}(34)= V_{\bar{q}\bar{q}}^{conf}(34)+V_{\bar{q}\bar{q}}^{OGE}(34)+V_{\bar{q}\bar{q}}^{ch}(34),\\
V_{Q\bar{q}}(ij)= V_{Q\bar{q}}^{conf}(ij)+V_{Q\bar{q}}^{OGE}(ij).
\end{array} \right.
\end{eqnarray}

For the heavy quark pair $(QQ)$, the confinement potential
$V_{QQ}^{conf}(12)$, which provides the non-perturbative QCD
effect in long distance, is taken as linear form
\begin{eqnarray}
V_{QQ}^{conf}(12)=-(\lambda_{1}^{c}\cdot\lambda_{2}^{c})(a_{12}r_{12}+a_{12}^{0})
.
\end{eqnarray}
$V_{QQ}^{OGE}(12)$ is the one-gluon-exchange interaction, whose
expression is
\begin{eqnarray}
V_{QQ}^{OGE}(12)=\frac{1}{4}g_{1}g_{2}(\lambda_{1}^{c}\cdot\lambda_{2}^{c})\
\left\{
\frac{1}{r_{12}}-\frac{\pi}{2}(\frac{1}{m_{1}^{2}}+\frac{1}{m_{2}^{2}}+
\frac{4(\bm\sigma_{1}\cdot \bm\sigma_{2})}{3m_{1}m_{2}})\delta(\bm
r_{12})\ \right\} .
\end{eqnarray}

For the light anti-quark pair ($\bar{q}\bar{q}$), we can obtain the
forms of $V_{\bar q\bar q}^{OGE}$ and $V_{\bar q\bar q}^{conf}$ by
replacing the $\lambda^c_i\cdot\lambda^c_j$ in Eq. (3) and Eq. (4)
by $\lambda^{c\ast}_i\cdot\lambda^{c\ast}_j$, and the chiral quark
interaction between two anti-quarks $V_{\bar q\bar q}^{ch}$ has the
same form (only $\lambda_{3}^{a}\cdot\lambda_{4}^{a}$ is replaced by
$\lambda_{3}^{a*}\cdot\lambda_{4}^{a*}$) as that of two light quarks
$V_{qq}^{ch}$. In the CSQ model, $V_{qq}^{ch}$ includes the scalar
boson exchanges and the pseudoscalar boson exchanges, thus
\begin{eqnarray}
V_{\bar q \bar q}^{ch}(34) = \sum_{a=0}^{8}V_{s_{a}}(\bm r_{34})+
\sum_{a=0}^{8}V_{ps_{a}}(\bm r_{34}) ,
\end{eqnarray}
and the expressions of these potentials are
\begin{eqnarray}
V_{s_{a}}(\bm r_{34}) =
-C(g_{ch},m_{s_{a}},\Lambda)(\lambda_{3}^{a*}\cdot\lambda_{4}^{a*})X_{1}(m_{s_{a}},\Lambda,r_{34})
,
\end{eqnarray}
\begin{eqnarray}
V_{ps_{a}}(\bm r_{34}) = C(g_{ch},m_{ps_{a}},\Lambda)
(\lambda_{3}^{a*}\cdot\lambda_{4}^{a*})
\frac{m_{ps_{a}}^{2}}{12m_{3}m_{4}}
X_{2}(m_{ps_{a}},\Lambda,r_{34})(\bm \sigma_{3}\cdot \bm
\sigma_{4})
,
\end{eqnarray}
where
\begin{eqnarray}
 C(g_{ch},m,\Lambda)=\frac{g_{ch}^{2}}{4\pi}\frac{\Lambda^{2}m}{\Lambda^{2}-m^{2}} ,
\end{eqnarray}
\begin{eqnarray}
X_{1}(m,\Lambda,r)=Y(mr)-\frac{\Lambda}{m}Y(\Lambda r) ,
\end{eqnarray}
\begin{eqnarray}
X_{2}(m,\Lambda,r)=Y(mr)-(\frac{\Lambda}{m})^{3}Y(\Lambda r) ,
\end{eqnarray}
\begin{eqnarray}
Y(x)=\frac{1}{x}e^{-x} ,
\end{eqnarray}
and $m_{s_{a}} (m_{ps_{a}})$ is the mass of the scalar
(pseudoscalar) meson.

The interactions of quark-antiquark pair ($Q\bar{q}$) include two
parts: direct interaction and annihilation part
\begin{eqnarray}
V_{Q\bar q}=V_{Q\bar q}^{dir}+V_{Q \bar q}^{ann} ,
\end{eqnarray}
\begin{eqnarray}
V_{Q \bar q}^{dir}=V_{Q\bar q}^{conf}+V_{Q\bar q}^{OGE} .
\end{eqnarray}
The contribution of annihilation part is neglected in this work,
and the expression of $V_{Q\bar q}^{dir}$ can be obtained from
$V_{Qq}^{dir}$. Concerning $V_{Q\bar q}^{conf}$ and $V_{Q \bar
q}^{OGE}$, the transformation from $V_{Qq}$ (the expressions of
$V_{Qq}^{conf}$ and $V_{Qq}^{OGE}$ are the same as Eqs.(3) and
(4)) to $V_{Q\bar{q}}$ is given by
$\lambda_{i}^{c}\cdot\lambda_{j}^{c}\longrightarrow
-\lambda_{i}^{c}\cdot\lambda_{j}^{c\ast}$, while for
$V_{Q\bar{q}}^{ch}$ the alteration is
$\lambda_{i}^{a}\cdot\lambda_{j}^{a}\longrightarrow
\lambda_{i}^{a}\cdot\lambda_{j}^{a\ast}$.

\subsection{Determination of the parameters}

The interaction parameters include the OGE coupling constant
$g_{i}$, the confinement strengths ($a_{ij},a_{ij}^{0}$), and the
chiral coupling constant $g_{ch}$. For the light quark pairs, the
parameters are taken from our previous works
\cite{hxzhang12,fhuang13}, which gave a satisfactory description for
the energies of the light baryon ground states, the nucleon-nucleon
scattering phase shifts the binding energy of the deuteron, where
the chiral coupling constant $g_{ch}$ is fixed by
\begin{eqnarray}
\frac{g_{ch}^{2}}{4\pi}=\frac{9}{25}\frac{g_{NN_{\pi}}^{2}}{4\pi}\frac{m_{u}^{2}}{M_{N}^{2}}
\end{eqnarray}
with $g_{NN_{\pi}}^{2}/4\pi=13.67$ taken as the experimental
value. The parameters for $cq$ or $cc$ quark pairs are taken from
Ref. \cite{hxzhang12}, which fitted the masses of $D, D^{*},
D_{s}, D_{s}^{*}, \eta_{c}, J/\Psi$ and $h_{1}(1p)$. Followed the
same method, the model parameters for $bq$ and $bb$ quark pairs
can be fixed by the masses of $B, B^{*}, B_{s}, B_{s}^{*},
\eta_{b}$ and $\Upsilon(1s)$. The complete set of the parameters
is shown in Table I, and the corresponding theoretical results for
the masses of $Q\bar{q}$ and $Q\bar{Q}$ mesons are shown in Table
II, from which we can see that the theoretical masses of mesons
are reasonably consistent with their experimental values,
respectively.

\begin{table}[htb]
\caption{\label{para} Model parameters for the systems with heavy
quarks. The masses of $u$ $(d)$ and $s$ quark: $m_{u,d}=313MeV$,
$m_{s}=470MeV$. The OGE coupling constant $g_{u}=0.886$.}
\begin{center}
\begin{tabular}{cccc|cccc}
\hline
 $m_{c}$ (MeV)& & 1430  & & &$m_{b}$ (MeV) & &4717 \\
 $g_c$& &  0.58 & & &$g_b$ & & 0.52 \\
 $a_{cu}$ (MeV/fm) & & 275 & & &$a_{bu}$ (MeV/fm) & &275 \\
 $a_{cu}^{0}$ (MeV)  & & -155.9& & &$a_{bu}^{0}$ (MeV) & &-141.1 \\
 $a_{cs}$ (MeV/fm)  &  & 275 & & &$a_{bs}$ (MeV/fm) & &275 \\
 $a_{cs}^{0}$ (MeV) &  & -124.7& & &$a_{bs}^{0}$ (MeV)  & &-112  \\
 $a_{cc}$ (MeV/fm)& &    275 &   &  &$a_{bb}$ (MeV/fm) & &275  \\
 $a_{cc}^{0}$ (MeV)& &-77.8 &  &  &$a_{cc}^{0}$ (MeV) & &-39.5 \\
 \hline
\end{tabular}
\end{center}
\end{table}

\begin{table}
\caption{The masses (MeV) of $Q\bar q$ and $Q\bar Q$ mesons.
Experimental data are taken from PDG.}
\begin{center}
\begin{tabular}{p{15mm}p{12mm}p{12mm}p{12mm}p{12mm}p{12mm}p{12mm}p{12mm}l}
\hline Mesons & $D$ & $D^{*}$ & $D_s$ & $D_s^{*}$ & $\eta_{c}$ & $J/\Psi$ & $h_{c}(1p)$ &\\
\hline
Exp.  & 1867.7 & 2008.9 & 1968.5 & 2112.4 & 2979.6 & 3096.9 & 3526.2 &\\
Theor. & 1888 & 2009 & 1969 & 2130 & 2990 & 3098 & 3568 &\\
\hline
Mesons & $B$ & $B^{*}$ & $B_s$ & $B_s^{*}$ & $\eta _b$ & $\Upsilon(1s)$ & &\\
\hline
Exp.  & 5279.2 & 5325 & 5369.6 & 5416.6 & 9300 & 9460.3 &  &\\
Theor. & 5288 & 5320 & 5317 & 5412 & 9404 & 9460 &   &\\
\hline
\end{tabular}
\end{center}
\end{table}

\subsection{Four-quark wave function}

For the description of a 4q state $(QQ\bar{q}\bar{q})$, the wave
function can be written as the following form
\begin{eqnarray}
\Psi(4q)=\psi_{4q}(0s^{4})[(QQ)^{I_{1}}_{S_{1},C_{1}}(\bar{q}\bar{q})^{I_{2}}_{S_{2},C_{2}}]^{I}_{S,C},
\end{eqnarray}
where $S$ represents spin, $I$ represents isospin, and $C$
represents color. $\psi_{4q}(0s^{4})$ is the orbital part and all
the four quarks are in (0s)-wave state;
$[(QQ)^{I_{1}}_{S_{1},C_{1}}(\bar{q}\bar{q})^{I_{2}}_{S_{2},C_{2}}]^{I}_{S,C}$
is the flavor-spin-color part. Since the wave function of
$(QQ)^{I_{1}}_{S_{1},C_{1}}$ and
$(\bar{q}\bar{q})^{I_{2}}_{S_{2},C_{2}}$ pairs should be
antisymmetric \cite{jvijande14}, the possible configurations of
$QQ\bar{q}\bar{q}$ are read as follows,
\begin{eqnarray}
\Psi_{A}=\psi_{4q}(0s^{4})[(QQ)_{0,6^{c}}(\bar{n}\bar{n})^{0}_{1,\bar{6}^{c}}]^{0}_{1,1^{c}},
\end{eqnarray}
\begin{eqnarray}
\Psi_{B}=\psi_{4q}(0s^{4})[(QQ)_{0,6^{c}}(\bar{n}\bar{n})^{1}_{0,\bar{6}^{c}}]^{1}_{0,1^{c}},
\end{eqnarray}
\begin{eqnarray}
\Psi_{C}=\psi_{4q}(0s^{4})[(QQ)_{1,\bar{3}^{c}}(\bar{n}\bar{n})^{0}_{0,3^{c}}]^{0}_{1,1^{c}},
\end{eqnarray}
\begin{eqnarray}
\Psi_{D_{J}}=\psi_{4q}(0s^{4})[(QQ)_{1,\bar{3}^{c}}(\bar{n}\bar{n})^{1}_{1,3^{c}}]^{1}_{J,1^{c}},
(J=0,1,2).
\end{eqnarray}
\begin{eqnarray}
\Psi_{E}=\psi_{4q}(0s^{4})[(QQ)_{0,6^{c}}(\bar{s}\bar{s})_{0,\bar{6}^{c}}]_{0,1^{c}},
\end{eqnarray}
\begin{eqnarray}
\Psi_{F_{J}}=\psi_{4q}(0s^{4})[(QQ)_{1,\bar{3}^{c}}(\bar{s}\bar{s})_{1,3^{c}}]_{J,1^{c}},
(J=0,1,2).
\end{eqnarray}
Thus for different quantum numbers $(J^{P};I)$, the configurations
of $QQ\bar{q}\bar{q}$ states can be written as
\begin{eqnarray}
|0^{+};0\rangle=\Psi_{E} ~and~ \Psi_{F_{0}} ,
\end{eqnarray}
\begin{eqnarray}
|0^{+};1\rangle=\Psi_{B} ~and~ \Psi_{D_{0}} ,
\end{eqnarray}
\begin{eqnarray}
|1^{+};0\rangle=\left\{\begin{array}{l}\Psi_{A} ~and~ \Psi_{C}\\
\Psi_{F_{1}} ,
\end{array} \right.
\end{eqnarray}
\begin{eqnarray}
|1^{+};1\rangle=\Psi_{D_{1}} ,
\end{eqnarray}
\begin{eqnarray}
|2^{+};0\rangle=\Psi_{F_{2}} ,
\end{eqnarray}
\begin{eqnarray}
|2^{+};1\rangle=\Psi_{D_{2}} .
\end{eqnarray}

We solve the Schrodinger equation of the 4q system by using the
variation method. The trail wave function is taken as an expression
of the 4q states with several different harmonic oscillator
frequencies $\omega_{i}$,
\begin{eqnarray}
\psi_{4q}(0s^{4})=\Sigma^{n}_{i}\alpha_{i}\phi_{4q}(b_{i}),
\end{eqnarray}
where $b_{i}^{2}=\frac{1}{m\omega_{i}}$. Then the energies of
these states can be obtained. For the 4q states with the same
$(J^{P};I)$, the configuration mixture has been considered in our
calculation.

\section{Results and discussions}

4q state will be stable under the strong interaction if their
total energy lies below all the thresholds of the possible and
allowed two-meson decay channels. Therefore, we use the quantity
\begin{eqnarray}
\Delta E=E(QQ\bar{q}\bar{q})-E_{m_{1}}(Q\bar{q})-E_{m_{2}}(Q\bar{q})
\end{eqnarray}
to discriminate the stable 4q states.

We calculated the energies of the 4q states with the quantum
numbers: $(J^{P};I)=(0^{+};0)$, $(0^{+};1)$, $(1^{+};0)$,
$(1^{+};1)$, $(2^{+};0)$, and$(2^{+};1)$. The results of
$cc\bar{q}\bar{q}$ and $bb\bar{q}\bar{q}$ are listed in table III
and IV, respectively. Our calculation predicts one bound state for
the $bb\bar{n}\bar{n}$ system with $(J^{P};I)=(1^{+};0)$ and no
bound state for the $cc\bar{q}\bar{q}$ system, both of which agree
with the conclusion that for a ratio $M/m\geq15$ ($M$ is the mass of
the heavy quark $c$ or $b$, $m$ is the mass of light quark $n$ or
$s$), a collective bound state appears, stable against spontaneous
dissociation \cite{szou15}. Meanwhile, we notice that the states
$cc\bar{n}\bar{n}$ and $bb\bar{n}\bar{n}$ with $(J^{P};I)=(1^{+};0)$
have the lowest energy than other $QQ\bar{q}\bar{q}$ states. For the
$(J^{P};I)=(1^{+};0)$ state, we calculated the energies of the
configurations $\Psi_{A}$ and $\Psi_{C}$ (Eqs. (16) and (18)) and
also the configuration mixing between $\Psi_{A}$ and $\Psi_{C}$ .
The results show that the energy of the configuration $\Psi_{C}$ is
lower and the mixture effect between $\Psi_{A}$ and $\Psi_{C}$ is
very small, thus the component $\Psi_{A}$ can be neglected and
component
$[(QQ)_{1,\bar{3}^{c}}(\bar{n}\bar{n})^{0}_{0,3^{c}}]^{0}_{1,1^{c}}$
is dominate. In $\Psi_{C}$, the light anti-quark pair
$(\bar{n}\bar{n})^{0}_{0,3^{c}}$  contributes a strong attractive
color magnetic force (CMF) which upholds Jaffe's di-quark model
\cite{rjaffe16}, at the same time the pseudoscalar meson exchange
interaction of $(\bar{n}\bar{n})_{0,3^{c}}^{0}$ is also strongly
attractive, thus $(\bar{n}\bar{n})_{0,3^{c}}^{0}$ can be
sufficiently attractive for forming a possible bound 4q state. On
the other hand, in the state
$[(QQ)_{1,\bar{3}^{c}}(\bar{n}\bar{n})^{0}_{0,3^{c}}]^{0}_{1,1^{c}}$,
the heavy quark pair $(QQ)_{1,\bar{3}^{c}}$ contributes a repulsive
CMF. When $Q=b$, the repulsive CMF is much weaker than that of the
attractive force contributed by the light anti-quark pair component
$(\bar{n}\bar{n})_{0,3^{c}}^{0}$, because of the large heavy quark
mass. Yet when $Q=c$, the repulsive CMF is not weak enough to
counteract the attractive force. That is why for the same
$(J^{P};I)=(1^{+};0)$ state, a bound state in $bb\bar{q}\bar{q}$
system can be found while there is no bound state in
$cc\bar{q}\bar{q}$ system.

\begin{table}
\caption{$E(cc\bar{q}\bar{q})$ and $\Delta E$ energies in MeV}
\begin{center}
\begin{tabular}{p{15mm}p{12mm}p{12mm}p{12mm}p{12mm}p{12mm}p{12mm}p{12mm}l}
\hline $(J^{P};I)$ & $(0^{+};0)$ & $(0^{+};1)$ & $(1^{+};0)$ & $(1^{+};0)$ & $(1^{+};1)$ & $(2^{+};0)$ & $(2^{+};1)$ &\\
\hline
  & $cc\bar{s}\bar{s}$ & $cc\bar{n}\bar{n}$ & $cc\bar{n}\bar{n}$ &$cc\bar{s}\bar{s}$ & $cc\bar{n}\bar{n}$ & $cc\bar{s}\bar{s}$ & $cc\bar{n}\bar{n}$&\\
 E & 4402 & 4161 & 4068 & 4462 & 4196 & 4504 & 4231 &\\
$\Delta E $& 142 & 143 & 50 & 363 & 299 & 244 & 213 &\\
\hline
\end{tabular}
\end{center}
\end{table}

\begin{table}
\caption{$E(bb\bar{q}\bar{q})$ and $\Delta E$ energies in MeV}
\begin{center}
\begin{tabular}{p{15mm}p{12mm}p{12mm}p{12mm}p{12mm}p{12mm}p{12mm}p{12mm}l}
\hline $(J^{P};I)$ & $(0^{+};0)$ & $(0^{+};1)$ & $(1^{+};0)$ & $(1^{+};0)$ & $(1^{+};1)$ & $(2^{+};0)$ & $(2^{+};1)$ &\\
\hline
  & $bb\bar{s}\bar{s}$ & $bb\bar{n}\bar{n}$ & $bb\bar{n}\bar{n}$ &$bb\bar{s}\bar{s}$ & $bb\bar{n}\bar{n}$ &$bb\bar{s}\bar{s}$ & $bb\bar{n}\bar{n}$&\\
 E & 10934 & 10695 & 10576 & 10945 & 10701 & 10959 & 10712 &\\
$\Delta E $& 192 & 119 & -32 & 216 & 93 & 135 & 72 &\\
\hline
\end{tabular}
\end{center}
\end{table}

Since CMF and the pseudoscalar meson exchange interaction can
contribute enough attractive force to form a bound 4q state, it is
interesting to find whether the existence of the bound state depends
on the contribution of CQI or not. Following the above procedure,
the parameters without considering CQI should be fixed firstly. Here
we briefly give the procedure for the parameter determination. The
one-gluon-exchange coupling constant $g_{u}$ is determined by the
mass split between $N$ and $\Delta$, and the confinement strengths
($a_{uu},a_{uu}^{0}$) are fixed by the masses of $N$ and $\Delta$.
($a_{cu},a_{cu}^{0}$) are fixed by the masses of $D$ and $D^{*}$,
while ($a_{bu},a_{bu}^{0}$) are fixed by the masses of $B$ and
$B^{*}$. Other parameters are taken from previous work
\cite{fhuang13}. The parameters without considering CQI are shown in
table V, and the theoretical results for the masses of $N$, $\Delta$
and $Q\bar{q}$ mesons without considering CQI are given in Table VI.

\begin{table}[htb]
\caption{\label{para} Model parameters without CQI. The quark masses
are as follows: $m_{u,d}=313MeV$, $m_{c}=1430MeV$, $m_{b}=4717MeV$.}
\begin{center}
\begin{tabular}{cccc}
\hline
 $g_{u}$ & & 0.945  & \\
 $a_{uu}^{0}$ (MeV)  & & -181.68& \\
 $a_{uu}$ (MeV/fm)  &  & 187.188 &  \\
 $a_{cu}^{0}$ (MeV) &  & -153.5&  \\
 $a_{cu}$ (MeV/fm)& &    275 & \\
 $a_{bu}^{0}$ (MeV)& &-138.8 &  \\
 $a_{bu}$ (MeV/fm)& &    275 &  \\
 \hline
\end{tabular}
\end{center}
\end{table}

\begin{table}
\caption{The masses (MeV) of $P$, $\Delta$ and $Q\bar{q}$ mesons.
Experimental data are taken from PDG.}
\begin{center}
\begin{tabular}{p{15mm}p{12mm}p{12mm}p{12mm}p{12mm}p{12mm}p{12mm}l}
\hline  & $P$ & $\Delta$ & $D$ & $D^{*}$ & $B$ & $B^{*}$ &\\
\hline
Exp.  & 939 & 1232 & 1967.7 & 2008.9 & 5279.2 & 5325 &\\
Theor. & 939 & 1237 & 1879& 2009 & 5287 & 5321 &\\
\hline
\end{tabular}
\end{center}
\end{table}

By using the variation method in section II.C, the masses of
$QQ\bar{q}\bar{q}$ states can be obtained. Here we only show the
numerical results for the states $cc\bar{n}\bar{n}$ and
$bb\bar{n}\bar{n}$ with $(J^{P};I)=(1^{+};0)$ in Table VII. From
Table VII, we can see that without considering CQI, the
configuration $bb\bar{n}\bar{n}$ with $(J^{P};I)=(1^{+};0)$ is
still a bound state and no bound state for $cc\bar{n}\bar{n}$
configuration. Meanwhile, it should be noticed that this bound
state is not as strong as that of the case with chiral quark
interactions.

\begin{table}
\caption{$E(bb\bar{n}\bar{n})$, $E(cc\bar{n}\bar{n})$ and $\Delta E$
energies in MeV}
\begin{center}
\begin{tabular}{ccc|cccl}
\hline & Without CQI & & & Considering CQI & &\\
\hline
& $cc\bar{n}\bar{n}$ & $bb\bar{n}\bar{n}$ & & $cc\bar{n}\bar{n}$ & $bb\bar{n}\bar{n}$ &\\
 E & 4078 & 10590 & & 4068 & 10576 &\\
$\Delta E $& 60  &-18& & 50 & -32 &\\
\hline
\end{tabular}
\end{center}
\end{table}

In order to see the differences between these two cases, we
calculate the contribution of each part for the configuration
$bb\bar{n}\bar{n}$ with $(J^{P};I)=(1^{+};0)$. We find that
without considering CQI, the strength of OGE interaction increases
and it contributes enough attractive force to form a bound state,
thus without CQI, $bb\bar{n}\bar{n}$ with $(J^{P};I)=(1^{+};0)$
can also be bound. In comparison, when considering CQI, the OGE
interaction and meson-exchange interaction both contribute
attractive forces, and the contribution of the pseudoscalar meson
exchange interaction is more favored to form a bound state. That
is why for the configuration $bb\bar{n}\bar{n}$ with
$(J^{P};I)=(1^{+};0)$, the bound state considering CQI is stronger
than without considering CQI.

\section{Summary}

In this work, we study the masses of $QQ\bar{q}\bar{q}$
heavy-light 4q system in the CSQ model, and try to see the
possibilities of the existence of bound states. Our calculation
shows the existence of only one bound four-quark state with
$(J^{P};I)=(1^{+};0)$ in the bottom sector and no bound state for
the charm sector. Besides, by comparing the masses of the same
four-quark state while considering CQI and without CQI, we find a
similar result that $bb\bar{n}\bar{n}$ with $(J^{P};I)=(1^{+};0)$
is bound but this binding energy is relatively weak.

\begin{acknowledgements}
This work is supported in part by the National Natural Science
Foundation of China, Grant No. 10475087 and No. 10775146
\end{acknowledgements}

\end{document}